**Retrieval-Augmented Generation in Biomedicine: A Survey of Technologies, Datasets, and Clinical Applications**

Jiawei He[1,2*], Boya Zhang[2], Hossein Rouhizadeh[2], Yingjian Chen[3], Rui Yang[4], Jin Lu[5], Xudong Chen[1], Nan Liu[4], Douglas Teodoro[2*]

[1]Hunan City University, Yiyang, Hunan, China.

[2]University of Geneva, Geneva, Geneva, Switzerland.

[3]Henan University, Kaifeng, Henan, China.

[4] National University of Singapore, Singapore, Singapore.

[5]LvZhiDao Information Technology Co., Ltd., Changsha, Hunan, China.

*Corresponding author(s). E-mail(s): joewellhe@gmail.com; douglas.teodoro@unige.ch;

## Abstract

Large language models (LLMs) in biomedicine face a fundamental conflict between static parameter knowledge and the dynamic nature of clinical evidence. Retrieval-Augmented Generation (RAG) addresses this by grounding generation in external data, yet it introduces new complexities in latency and architecture. This survey synthesizes the biomedical RAG landscape (2020–2025), classifying systems into naive, advanced, and modular paradigms. Beyond a technological taxonomy, we formalize the biomedical RAG trilemma, identifying the inherent trade-offs between reasoning depth, inference latency, and data privacy that constrain current clinical deployment. We analyze how recent agentic workflows enhance diagnostic reasoning but risk prohibitive latency, and how privacy constraints dictate the choice between powerful cloud-based models and local deployment. Finally, we outline the alignment gap in multimodal RAG and propose future directions for self-correcting, verifiable clinical agents.

# 1 Introduction

While large language models (LLMs) demonstrate reasoning capabilities essential for clinical decision support, their reliance on static parametric memory creates a fundamental misalignment with the dynamic nature of biomedical evidence [1-4]. In clinical environments where guidelines evolve daily, the static knowledge limitation of standard LLMs poses severe risks to patient safety, manifesting as hallucinations that are plausible yet factually incorrect [5].

Retrieval augmented generation (RAG) aims to resolve this by decoupling reasoning (the generative model) from knowledge storage (the retrieval index) [6]. This architectural separation transforms the LLM from a memorizer into a reasoner, allowing for the real-time integration of up-to-date clinical data and knowledge without expensive retraining, while ensuring that generated claims can be grounded in verifiable citations [1-2].

However, standard plug-and-play RAG pipelines frequently fail in biomedical settings due to the lexical-semantic gap. Unlike general domains, biomedical retrieval must navigate complex ontologies (e.g., UMLS) where keyword matching fails to align lay terminology (e.g., *heart attack*) with clinical nomenclature (e.g., *acute myocardial infarction*), necessitating specialized dense retrieval and re-ranking architectures [7-9].

Existing reviews have predominantly focused on the *training* paradigms of medical LLMs (e.g., Liu *et al.* [10] and He *et al.* [2]), general-domain RAG taxonomies (e.g., Fan *et al.* [1] and Gao *et al.* [11]), or multimodal LLMs (Xiao *et al.* [4]). These works largely treat the retrieval component as a black box, overlooking specificities of sparse and dense retrieval in high-stakes clinical workflows. Furthermore, they fail to address the latency vs. accuracy dilemma inherent to deploying modular RAG systems in real-time hospital environments.

Our work differs from existing surveys by moving beyond enumeration to provide a synthesis of the RAG technologies in biomedicine. The key contributions of this survey are:

1. A novel taxonomy and evolutionary timeline: We classify biomedical RAG systems into distinct architectural paradigms (naive, advanced, and modular) and provide a timeline chart showcasing the co-evolution of domain-specific retrievers (e.g., MedCPT) and generative models.
2. Formalization of the biomedical RAG trilemma: We synthesize the architectural trade-offs into a unified theoretical framework, the biomedical RAG trilemma, analyzing how current systems must balance reasoning depth, inference latency, and data privacy.
3. Critical assessment of data and reliability: We analyze the implications of dataset biases, including Anglocentric and demographic imbalances in sources like MIMIC-IV, and discuss essential methodologies for measuring confidence and quantifying uncertainty in clinical outputs.

4. Comparative clinical analysis: We present concrete case studies (e.g., rare disease identification) that explicitly contrast conventional NLP performance with RAG-enhanced workflows, demonstrating the tangible benefits in accuracy and explainability.

## 2 Scope and Survey Taxonomy

To provide a comprehensive synthesis of the field, we adopted a systematic mapping approach. We queried major repositories (PubMed, arXiv, ACL Anthology) using the Boolean logic: (RAG OR 'retrieval-augmented') AND (biomedical OR clinical OR medical). To distinguish our analysis from general LLM surveys, we selected studies that i) explicitly couple a non-parametric retriever (vector database, knowledge graph (KG), or search API) with a parametric generator; ii) address specific biomedical challenges (e.g., ontology alignment, citations) rather than generic question-answering (QA); and iii) prioritize peer-reviewed works or open-source frameworks (e.g., MedCPT [12], Self-BioRAG [13]) that allow for architectural deconstruction. We excluded works relying solely on prompt engineering or long-context windows, as these lack the retrieval-based reasoning mechanism that is the focus of this review.

## 3 Problem Formulation

We formalize Biomedical RAG as the maximization of the generation probability $P(y|x)$ for a clinical query $x$, conditioned on a latent set of retrieved documents $Z$.

### 3.1 Mathematical Framework

The generation process is defined as:

$$P(y|x) = \sum_{z \in Z} P_{\theta}(y|x,z) P_{\phi}(z|x),$$

where $P_{\phi}(z|x)$ represents the retrieval likelihood (optimized via dense/sparse alignment) and $P_{\theta}(y|x,z)$ represents the generative reasoning capability. In modular RAG (Section 5.3), this expands to include a policy π where the model iteratively decides to retrieve, reason, or terminate.

### 3.2 General Execution Framework

To standardize the comparison of architectures discussed in Section 4, we define the canonical biomedical RAG workflow in Algorithm 1.

Algorithm 1: Canonical biomedical RAG workflow

**Input** clinical query $x$, document index $D$ , retriever $R$, generator $G$, threshold $\tau$

**Output** clinical response $y$, citations $C$

1 **Step 1: retrieval**

2 $Z_{raw} \leftarrow R(x, D)$ *Retrieve top-k candidates (sparse/dense)*

3 **Step 2: refinement (advanced RAG)**

4 $Z_{ranked} \leftarrow \text{Rerank}(x, Z_{raw})$ *Filter noise via cross-encoder*

5 **Step 3: generation and verification (modular RAG)**

| | |
|---|---|
| 6 | IF confidence$(G, c) < \tau$ |
| 7 | $y \leftarrow$ self-correct$(x, Z_{ranked})$ *Iterative refinement (e.g., Self−BioRAG)* |
| 8 | ELSE |
| 9 | $y \leftarrow G(x, Z_{ranked})$ |
| 10 | RETURN $y$ |

# 4 Biomedical RAG Architectures and Taxonomy

We classify architectures based on how they approximate and optimize this equation: *naive* RAG (linear approximation), *advanced* RAG (optimization of $P_\phi$), and m*odular* RAG (introduction of a policy $\pi$ to navigate $Z$). These paradigms reflect a shift from simple fact retrieval to complex clinical reasoning. While naive systems suffice for surface-level queries, the field is trending toward modular architectures to handle multi-hop reasoning required for more complex clinical scenarios, such as differential diagnosis. A comparative analysis of these biomedical RAG approaches is shown in **Table 1**.

The foundational architecture, often termed *naive* RAG, follows a traditional retrieve-read process. It assumes a single-hop linear approximation where the top-$k$ documents retrieved by a sparse or dense index are concatenated directly into the context window. This assumes that the retrieval score $S_R(q, d)$ is a perfect proxy for semantic relevance. As illustrated in Figure 1, a query $q$ is passed to a retriever $R(q, K)$, which identifies a set of documents $D = \{d_1, \ldots, d_n\}$ from a knowledge source $K$. These documents are concatenated directly with the query and fed into the LLM to generate a response. This architecture might suffer from grounded hallucination due to lexical-semantic dissonance. For example, if a retriever fetches guidelines on *viral pneumonia* for a *bacterial pneumonia* query due to keyword overlap or semantic relatedness, the LLM might generate a clinically fluent but incorrect treatment plan. Despite this risk, naive RAG remains the standard where latency is prioritized over complex reasoning.

*Advanced* RAG introduces a post-retrieval optimization stage: re-ranking. While the initial retrieval relies on computationally efficient bi-encoders ($O(N)$ via approximate nearest neighbor search), advanced RAG employs cross-encoders ($O(K)$) to perform full self-attention over the query-document pairs. While this might significantly improve Precision@$K$ by filtering or downgrading irrelevant context, it introduces a latency bottleneck (often > 500ms) [14]. Thus, while advanced RAG might be preferred for physician-facing clinical decision support systems (CDSS), where accuracy is non-negotiable, response time might be prohibitive for real-time conversational agents.

Recent developments have led to *modular* RAG, which creates flexible, non-linear workflows suited for complex biomedical reasoning. Unlike the static pipelines of naive and advanced RAG, modular systems employ iterative retrieval or routing mechanisms. In this sense, modular systems represent a shift toward agentic RAG. Systems, such as Self-BioRAG [13] and GeneGPT [15], introduce a routing policy $\pi(a|s)$, where the model iteratively decides an action $a$, such as whether to *retrieve* from a vector store, *compute* using an external tool (e.g., a genomic calculator), or *generate* a final answer, conditioned on the current context of the system $s$. While this allows for 'System 2' reasoning (deliberative thought) required for differential diagnosis, it introduces a latency penalty (refer to the trilemma in Section 8), subjecting the system to non-deterministic execution times.

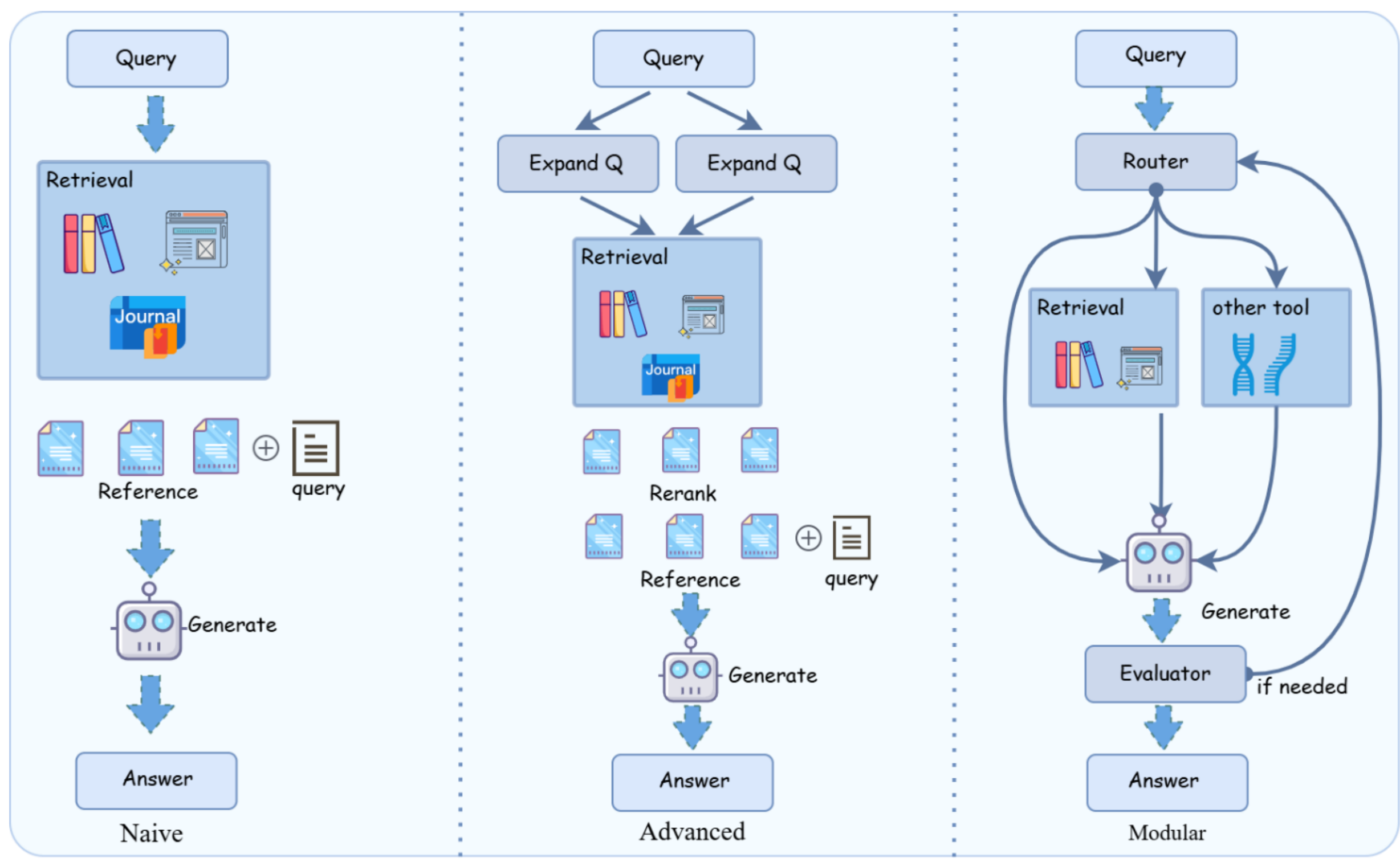

**Figure 1**: Basic biomedical RAG framework.

**Table 1:** Comparative analysis of biomedical RAG paradigms (naive, advanced, modular).

| Feature | Naive RAG | Advanced RAG | Modular RAG |
|---|---|---|---|
| Mathematical Model | linear approximation | optimized $P(z\|x)$ (re-ranking) | iterative policy $\pi(a\|s)$ |
| Inference latency | low (< 2s) [16] | medium (2-5s) [17] | high (> 10s) [17] |
| Clinical failure mode | hallucination via irrelevant documents | latency timeout | infinite loops / API failures |
| Best use case | patient education / FAQ | CDSS | rare disease diagnosis / genomics |

## 5 Technology

We trace the evolution of biomedical retrievers from lexical matching (2020) to the inclusion of instruction-tuned semantic alignment (2025). This shift was necessitated by the failure of keyword-based systems to capture the nuance of clinical queries (e.g., *heart failure* implying *cardiac insufficiency*), leading to the adoption of complementary dense embeddings that align vector space with medical ontologies.

### 5.1 Retriever

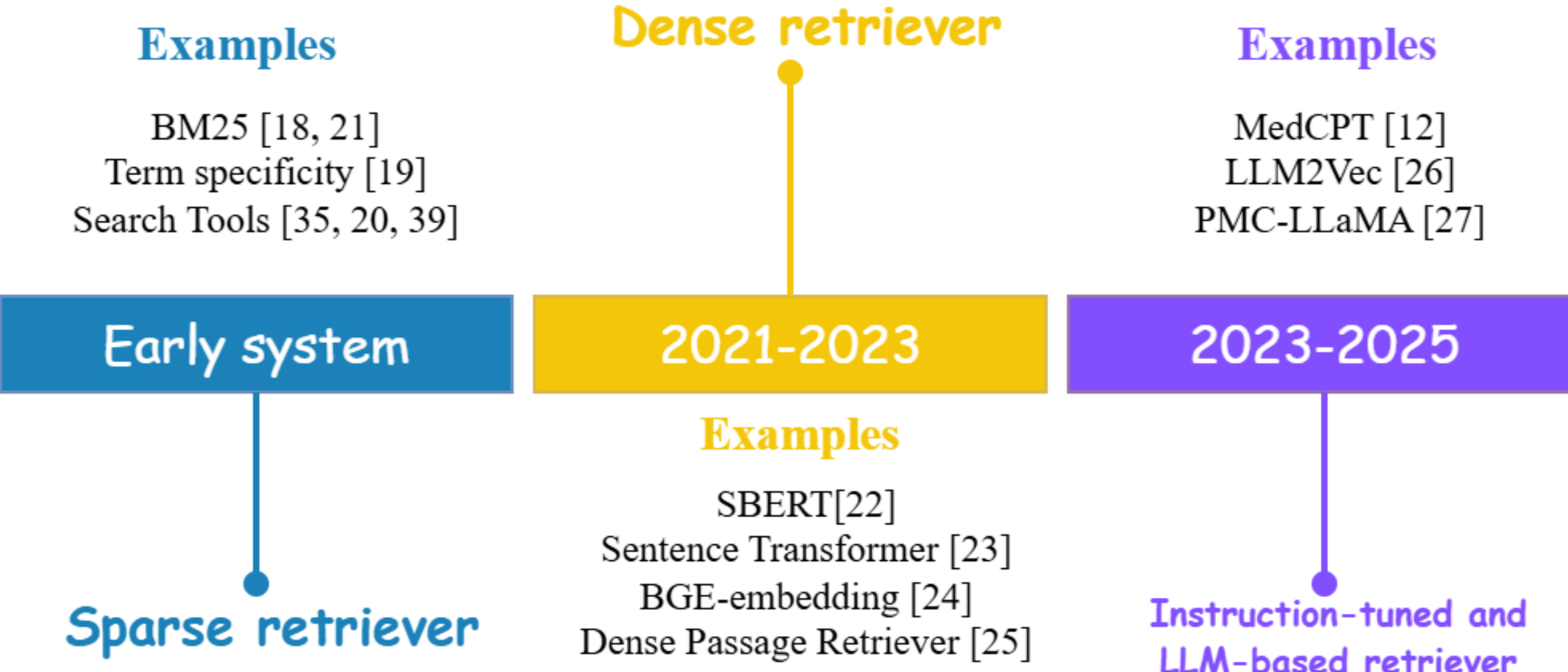

**Figure 2**: The evolutionary timeline of retrievers.

Retrievers function as the non-parametric memory access mechanism for the RAG system [11]. As illustrated by the evolutionary timeline in Figure 2, biomedical retrieval strategies have shifted significantly over the surveyed period (2020–2025). Early systems predominantly relied on sparse retrieval (BM25) [17-20] for precise keyword matching, e.g., of drug names and symptoms. The introduction of BERT-based dense retrievers (2021–2022) enabled semantic matching [21-24], while the most recent wave (2023–2025) utilizes instruction-tuned retrievers like MedCPT and LLM-based embeddings to align retrieval directly with clinical reasoning tasks [12, 25, 26].

#### 5.1.1 Sparse Retrieval

Sparse retrieval, exemplified by algorithms like BM25 [18] and TF-IDF [19], operates on the principle of lexical matching, where documents are indexed based on term occurrences rather than semantic meaning. While conceptually simple, as shown in Table 2, these methods remain indispensable in biomedicine due to the *lexical gap*. Neural models frequently exhibit semantic drift, associating clinically distinct antonyms (e.g., *hypoglycemia* vs. *hyperglycemia*) due to their shared vector space neighborhoods. In contrast, sparse retrieval enforces exact matching, ensuring that specific alphanumeric variants (e.g., ICD-10 E11.9) are retrieved accurately, preventing dangerous hallucinations in downstream generation. Thus, despite the rise of neural methods, sparse retrieval remains a competitive baseline [19, 20] and an essential component in hybrid systems [27-30], combining sparse and dense retrieval techniques. Recent studies consistently demonstrate that fusing BM25 with dense vectors enhances generalization, effectively handling the lexical gap where precise medical nomenclature does not align with semantic clusters [31-33].

**Table 2**: Sparse retrieval methods in biomedical RAG systems.

| Method name | Retrieval method | Task | Year |
| --- | --- | --- | --- |
| Clinfo.ai [35] | Entrez API [36] | medical QA | 2023 |
| MEDRAG toolkit [7] | Hybrid (BM25, SPECTER, Contriever, MedCPT) | medical QA | 2024 |
| SeRTS [33] | Self-Rewarding Tree Search based BM25 | biomedical QA | 2024 |
| Dual RAG System [28] | Hybrid (dense+BM25 with language-specific tokenizers) | diabetes management | 2024 |
| MedExpQA [29] | BM25 and MedCPT | medical QA | 2024 |
| CliniqIR [30] | BM25 and MedCPT | diagnostic decision support | 2024 |
| VAIV Bio-Discovery [31] | Hybrid (dense+BM25) | biomedical knowledge discovery | 2024 |
| Shuangjia et. al [37] | *Azure AI Search* with keyword matching | genetic variant annotation | 2025 |
| Eun Jeong et. al [21] | BM25 | medical QA | 2025 |
| Sudeshna et. al [20] | Whoosh [38] | medical QA | 2025 |
| LITURAt [39] | Entrez API [36] | scientific data analysis | 2025 |

### 5.1.2 Dense Retrieval

Dense retrievers encode queries and documents into a continuous vector space where semantic similarity is measured via cosine distance [40]. The most effective biomedical retrievers achieve domain alignment through two primary adaptation strategies: continued pre-training on biomedical corpora (e.g., BMRETRIEVER [32]) and contrastive fine-tuning (e.g., MedCPT [12]). This training objective ensures that a query embedding $q$ is closer to its relevant positive document $d^+$ than to a set of negatives $D^-$, mitigating the semantic disparities between general and biomedical language domains.

We classify dense retrievers into a hierarchy of utility vs. privacy (Table 3):

1. Type 1 - General PLMs: PLMs, such as BERT, often fail to capture medical nuances [21-24, 38-52].
2. Type 2 - Commercial APIs: While offering superior semantic reasoning (e.g., OpenAI), they pose significant data residency risks (GDPR/HIPAA), limiting their use to non-patient-identifiable data [27, 53-64].
3. Type 3 - Domain-adapted retrievers: Models like MedCPT balance privacy and performance but require substantial fine-tuning resources [13, 65-72].
4. Type 4 - LLM-based embeddings: The emerging state-of-the-art, allowing for instruction-based retrieval (e.g., retrieve only *contraindications*), though at a higher computational cost [25, 73, 74].

**Table 3**: Classification of dense retriever types in biomedical RAG systems. A more detailed taxonomy is provided in Appendix Table A1.

| Type | Definition | Representative example | Trade-off (Trilemma) |
|---|---|---|---|
| 1 - General PLMs | off-the-shelf BERT models without domain tuning. | AskFDALabel [78] | high availability / low accuracy |
| 2 - Commercial APIs | proprietary endpoints (e.g., OpenAI) with superior semantic alignment. | DRAGON-AI [56] | high accuracy / low privacy |
| 3 - Domain-adapted | open-source models fine-tuned on biomedical corpora (PubMed). | MedCPT [12] | balanced privacy and accuracy / high training cost |
| 4 - LLM-based | instruction-tuned embeddings allowing task-specific queries. | BiomedRAG [5] | high accuracy / high latency |

Implementation of dense retrievers relies on vector stores like Faiss [79] and Chroma [80] (Table 4). However, for clinical deployment, the choice of infrastructure is dictated by privacy; local instances are preferred over managed cloud solutions to ensure PHI (protected health information) remains on-premise. While cloud-native stores offer auto-scaling, hospital on-premise introduces an infrastructure and maintenance overhead that is often underestimated in research papers.

**Table 4**: Common vector stores in biomedical RAG

| Store | Description |
| --- | --- |
| Chroma [80] | AI-native open-source vector database with built-in functionality |
| Faiss [79] | Library for efficient similarity search and clustering of dense vectors |
| Pinecone [81] | Production-ready vector database for similarity search at scale |
| Weaviate [82] | Graph-based vector search engine with semantic capabilities |
| HNSW [51] | Hierarchical navigable small world graphs for approximate nearest neighbor search |

**The Precision-Recall Trade-off.** The analysis of current retrieval paradigms reveals a tension between *semantic recall* and *lexical precision*. While dense retrievers (e.g., MedCPT ) excel at mapping symptoms to diagnoses, they frequently fail on exact alphanumeric matches (e.g., ICD-10 codes). Conversely, sparse retrieval ensures lexical safety but misses semantic nuance. This necessitates the hybrid architectures seen in recent clinical systems, where the computational cost of dual-indexing is accepted as the price for patient safety.

## 5.2 Reranker

Rerankers in biomedical RAG systems serve as intermediary components that refine and prioritize the relevance of initially retrieved documents before they are processed by the generative models [83]. Positioned between retrievers and generators, rerankers apply sophisticated algorithms to re-score and re-order candidate documents based on their contextual relevance to the query, effectively functioning as a second-stage filtering mechanism that enhances the precision of information ultimately provided to the generation component.

Biomedical RAG rerankers can be categorized into distinct types based on their underlying methodologies:

- **Statistical rerankers** apply traditional information retrieval techniques; for instance, CliniqIR [30] and MEDRAG [73] both implement Reciprocal Rank Fusion (RRF) to combine multiple retrieval signals, while DRAGON-AI [56] employs Maximal Marginal Relevance (MMR) to balance relevance with diversity.
- **Content-based rerankers**, in contrast, focus on content similarity; notably, WeiseEule [73] introduces a keyword frequency-based ranking system specifically designed for biomedical literature retrieval.
- **Model-based rerankers** employ transformer architectures to perform contextual relevance assessment, effectively functioning as a domain adaptation strategy. For instance, MedCPT [12] trains a specialized cross-encoder on PubMed search logs, while BiomedRAG [5] implements a domain-specific chunk scorer to prioritize evidence containing causal medical relationships. These specialized mechanisms capture subtle signals that general embedding models often miss.

- **LLM-based rerankers** leverage foundation models' reasoning capabilities. For example, GNQA [50] employs GPT-3.5/4/4o as zero-shot binary classifiers for relevance determination, and LmRaC [84] implements an LLM-based paragraph usefulness assessment with a minimum threshold score of 7 (on a 10-point scale).
- **Hybrid approaches** combine multiple strategies to enhance performance. For example, Clinfo.ai [35] combines LLM classification with BM25 to re-rank retrieved articles.

In practice, we observe a standard funnel architecture in clinical deployment: a computationally cheap retriever (bi-encoder/BM25) fetches 50-100 candidates, which are then re-scored by an expensive cross-encoder (model-based). While LLM-based reranking (e.g., GPT-4+) offers the highest accuracy, its latency (>1s per query) [85] hinders its application in real-time patient/physician-facing interfaces.

**The Accuracy-Latency Bottleneck.** In the context of the biomedical trilemma, rerankers represent the primary constraint on inference latency. While cross-encoders (model-based) maximize reasoning depth by effectively filtering noise, they increase latency by an order of magnitude compared to bi-encoders. This forces a hard choice in architectural design: real-time conversational agents must often bypass sophisticated reranking (sacrificing precision), whereas asynchronous CDSS can afford the computational cost for maximum safety.

## 5.3 Generation

The generation component represents the last element in biomedical RAG systems, responsible for synthesizing retrieved information into human-like responses that address user queries. In biomedical contexts, this process requires balancing natural language fluency with domain-specific accuracy, as generated content often informs healthcare decisions [86]. This section examines various generation models employed in recent biomedical RAG studies.

Current generation models employed in biomedical RAG systems can be categorized into three distinct classes based on their development approach and specialization: general-domain open-source models, commercial proprietary models, and biomedical-specialized models. These categories present different characteristics in terms of accessibility, customization potential, and domain-specific performance. Table 5 showcases representative models from each category and highlights recent studies that utilize these models (a more detailed taxonomy of the surveyed work is provided in Appendix Table A2). It is important to note that our analysis of biomedical-specialized LLMs focuses specifically on models trained directly on biomedical or healthcare corpora, deliberately excluding applications developed through prompt engineering or chain-of-thought (CoT) methodologies.

As shown in Table 5, the majority of LLMs employ decoder-only architectures, with T5 being a notable exception. This architectural preference stems from several inherent advantages for generative tasks, particularly the natural suitability for next token prediction and significantly improved parameter efficiency [87]. Among open-source LLMs and Llama 2 [88] is the most popular, with 15 biomedical applications built upon it. LLaMa 3 [89], Mistral 7B [90], and T5 [91] are also widely used. These open-source LLMs have various sizes, from 0.06B (T5) to 671B (Deepseek R1), providing flexible options for building Biomedical RAG systems. Recently, several powerful general open-source LLMs have emerged, including, Mistral Small 3.1 [92] and Gemma 3 [93]. Although we have not found these newer models being used for biomedical RAG systems in our survey, we believe they show promise due to their outstanding performance in general domain tasks.

For commercial LLMs, the ChatGPT series APIs maintain a dominant position, presumably attributable to their superior performance. Nevertheless, several studies have demonstrated that Claude and Gemini achieve comparable efficacy when implemented in biomedical RAG applications [55, 91, 92], thereby offering viable alternatives to ChatGPT. Furthermore, it is noteworthy that both DeepSeek and Mistral have established dual accessibility paradigms: providing open-source parameters for their base models while simultaneously offering commercial API services.

Biomedical specialized LLMs are developed through domain adaptation strategies, including continued pre-training on medical corpora (e.g., BioMistral [96], ChatENT [61]), supervised fine-tuning (e.g., MEDGENIE [6]), and parameter-efficient approaches like low-rank adaptation (LoRA) used in AskFDALabel [78]. Despite these specialization efforts, recent studies [72, 94, 95] demonstrate that within RAG frameworks, these models do not consistently outperform general open-source alternatives. We attribute this to the reasoning-knowledge decoupling effect. In RAG, the *knowledge* comes from the retrieved documents, so the primary role of an LLM is *reasoning* and *synthesis*. Massive generalist models possess superior reasoning capabilities derived from diverse training data. In contrast, smaller, domain-specific models often sacrifice reasoning power for vocabulary memorization, which is counterproductive in a RAG architecture.

**Table 5**: Taxonomic classification of LLMs for embedding applications.

| Tier | Model class | Representative models | Reasoning capability | Privacy Safety | Ideal Use Case |
|---|---|---|---|---|---|
| 1 - Cloud SOTA | proprietary API | GPT-4o, Claude 3.5 Sonnet, Gemini 1.5 Pro | very high (system 2) | low (data residency risk) | differential diagnosis, complex multi-step reasoning |
| 2 - Local generalist | open-weights (large) | Llama 3 (70B), Qwen 2.5 (72B), DeepSeek-R1 | high | high (on-premise) | hospital-side CDSS, summarization of PHI |
| 3 - Local specialist | domain-adapted (small) | BioMistral (7B), PMC-LLaMA (13B) | medium | high (on-premise) | edge devices, specific clinical and administrative tasks (e.g., note structuring) |
| 4 - Legacy | encoder-decoder | T5, BART | low (pattern matching) | high | text normalization, simple QA |

**The Reasoning-Privacy Conflict.** Our review identifies a divergence in generator selection driven by the privacy vs. reasoning axis of the biomedical trilemma. While commercial APIs (GPT-4) offer superior reasoning for complex differentials, they pose data residency risks. Conversely, open-source models (Llama 3, Mistral) allow for secure, on-premise deployment but often lack the "System 2" reasoning depth required in complex cases (e.g., rare disease diagnosis) without extensive fine-tuning.

## 5.4 Knowledge Graph-based RAG

While vector retrieval excels at capturing implicit semantic relationships (e.g., synonyms), it struggles with explicit multi-hop reasoning (e.g., drug A treats disease B, which is caused by gene C). KG retrieval resolves this by traversing structured edges [99]. On the negative side, it introduces a high maintenance cost. Unlike vector stores, which ingest unstructured text instantly, KGs require ontology alignment, making them brittle in the face of rapidly emerging medical concepts (e.g., COVID-19 and its subsequent variants).

Table 6 summarizes recent RAG implementations (2024–2025) using KG. We observe a divergence in application logic: while early systems focused on static fact retrieval (QA) [69, 97], recent frameworks leverage KGs for predictive reasoning (e.g., diagnosis prediction) [101] and hypothesis generation [102].

**Table 6**: Recent biomedical RAG studies based on KGs.

| Method | Task | KG Used | Year |
|---|---|---|---|
| KG-RAG [72] | biomedical text generation, medical question answering | SPOKE biomedical KG | 2024 |
| HEALIE [103] | personalized medical content generation | HEALIE KG (customized) | 2024 |
| KRAGEN [104] | Biomedical problem solving | Alzheimer's KG (AlzKB) | 2024 |
| Ascle [105] | various medical text generation tasks | UMLS | 2024 |
| DALK [106] | Alzheimer's QA | Alzheimer's KG (AlzKB) | 2024 |
| Gilbert *et al.* [107] | medical information structuring and interlinking | Not specified | 2024 |
| NetMe 2.0 [100] | biomedical knowledge extraction | BKGs constructed using OntoTagMe and Wikidata | 2024 |
| NEKO [108] | knowledge mining in synthetic biology | PubMed-derived KGs | 2025 |
| DR.KNOWS [101] | diagnosis prediction from EHRs | UMLS | 2025 |
| ESCARGOT [102] | biomedical reasoning and knowledge retrieval | Alzheimer's KG (AlzKB) | 2025 |

As shown in Figure 3, the integration of knowledge graphs with LLMs in biomedical RAG applications can be analyzed through four key components:

- **Biomedical KG construction**: We classify approaches into static authority vs. dynamic extraction. Systems like KG-RAG [72] utilize pre-curated, authoritative graphs (e.g., SPOKE [109]) to ensure high precision but suffer from stale knowledge. In contrast, on-the-fly systems like NetMe 2.0 [100] and NEKO [108] construct dynamic subgraphs from retrieved literature using entity linking. While this resolves the knowledge currency issue, it introduces extraction noise, potentially polluting the reasoning chain with incorrect triples.
- **Query structuration:** This stage addresses the translation of natural language into formal query languages (e.g., Cypher [102]). A limitation here is the alignment gap: LLMs often generate syntactically correct but semantically invalid queries when mapped to rigid schemas like UMLS [101]. Frameworks like ESCARGOT attempt to mitigate this via graph-of-thoughts (GoT) [102], while KRAGEN bypasses formal query generation entirely by vectorizing the graph nodes for semantic search, trading query expressivity for robustness [104].

- **Subgraph extraction:** Retrieving a full *k*-hop neighborhood around an entity often leads to context explosion, overwhelming the context window of LLMs with irrelevant nodes. To manage this signal-to-noise ratio, recent systems employ path-ranking algorithms [98, 100, 101]. For instance, DR.KNOWS [101] utilizes attention mechanisms to score and prune paths, isolating only the diagnostic pathways relevant to the patient's EHR data rather than the entire disease ontology.
- **Prompt integration:** The final challenge lies in graph linearization, i.e., converting structured triples into natural language prompts without losing topological information. While KG-RAG [72] implements token-optimized linearization to reduce costs, newer approaches like NetMe 2.0 [100] integrate graph data as soft logic constraints, guiding the LLM-generated content to remain faithful to the retrieved topology.

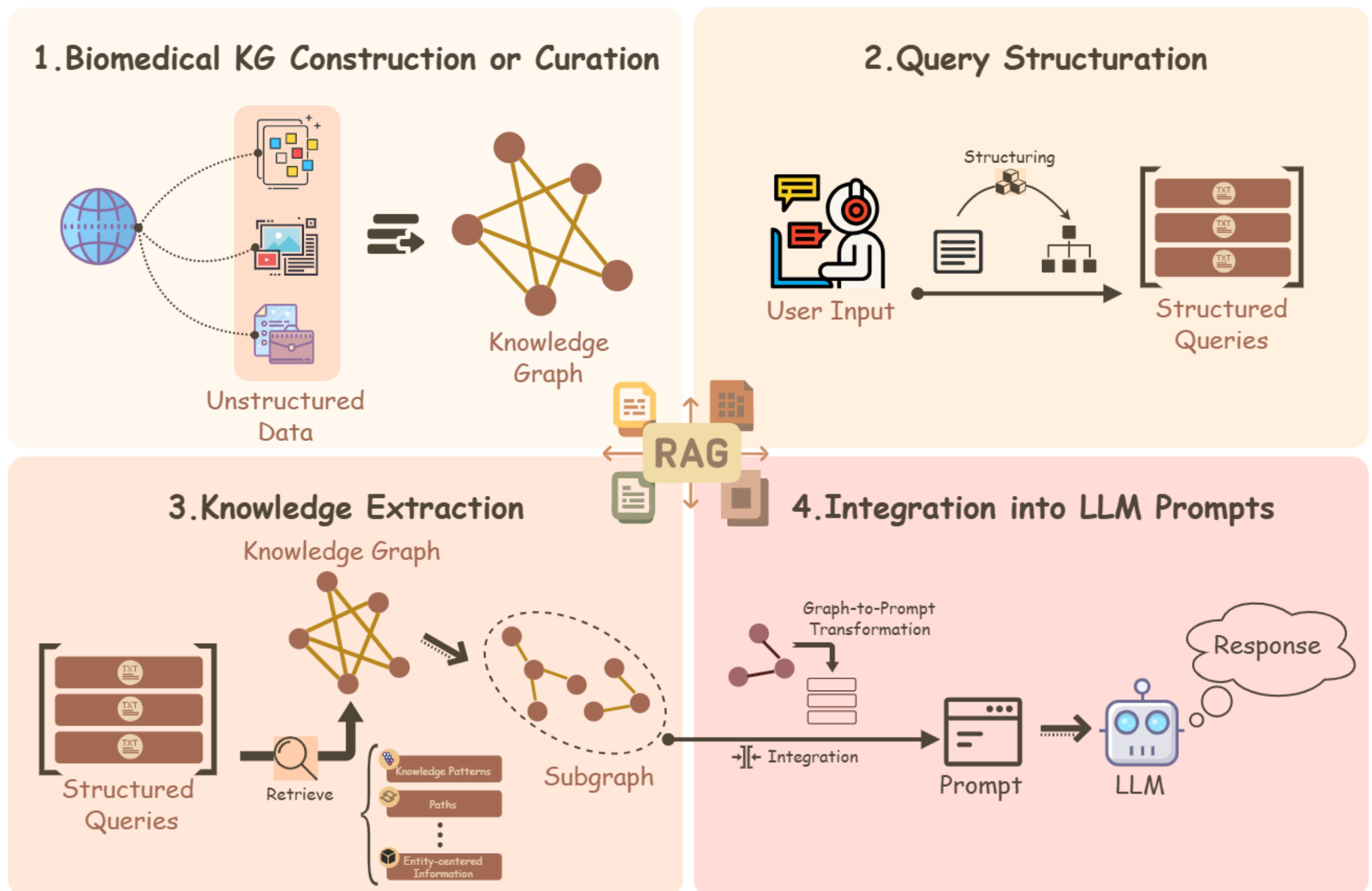

**Figure 3**: The framework of KG-based biomedical RAG: 1. Construction, 2. Structuration, 3. Extraction, and 4. Integration.

## 5.5 Multimodal RAG

Multimodal biomedical RAG addresses the semantic gap between high-dimensional pixel/sensor data and low-dimensional clinical concepts [110]. Unlike general-domain models trained on billion-scale image-text pairs (e.g., CLIP [111]), biomedical systems operate in a data-scarce regime where aligned modalities (e.g., MRI + caption [109-113]) are expensive to curate. Consequently, the challenge shifts from simple retrieval to cross-modal alignment, requiring specialized encoders to map disparate inputs (e.g., histopathology patches [117]) into a shared semantic vector space compatible with text embeddings. As shown in Table 7, multimodal RAG systems have been investigated in several scenarios in biomedicine, mostly combining diverse medical imaging and textual data.

**Table 7**: Multimodal biomedical RAG systems.

| Model | Task | Modalities | Year |
|---|---|---|---|
| Ranjit *et al.* [113] | chest X-Ray report generation | chest X-ray and text | 2023 |
| FactMM-RAG [112] | radiology report generation | X-Ray and text | 2024 |
| MMed-RAG [114] | medical VQA & report generation | medical images and text | 2024 |
| MEAG [118] | amblyopia diagnosis | eye tracking and text | 2024 |
| Tozuka *et al.* [119] | lung cancer staging | CT findings and TNM classifications | 2024 |
| Raminedi *et al.* [120] | radiology report generation | X-ray images and text | 2024 |
| RULE [115] | medical VQA & report generation | medical images and text | 2024 |
| Thetbanthad *et al.* [121] | prescription label identification | image (labels) and text | 2025 |
| Hu *et al.* [122] | pathology report generation | whole slide images and text | 2025 |
| STREAM [116] | chest X-ray report generation | chest X-ray and text | 2025 |

Figure 4 illustrates the architectural workflow of a multimodal biomedical RAG system. The key distinction between multimodal biomedical RAG and traditional biomedical RAG lies in data input diversity and processing architecture. While traditional RAG systems primarily process textual queries to retrieve text-based medical knowledge, multimodal RAG incorporates and processes non-textual data (images, sensor readings, etc.) alongside text, requiring specialized encoders for each modality to meaningfully integrate these diverse inputs [110].

The basic framework of multimodal biomedical RAG typically consists of: (1) *modality-specific encoders* (e.g., Vision Transformers for images, specialized encoders for time-series data [123]);

(2) *knowledge retrieval components* that access domain-specific databases based on multimodal query representations; and (3) *generation modules* that synthesize comprehensive outputs informed by both the multimodal inputs and retrieved knowledge.

We categorize encoding strategies into *symbolic conversion* vs. *neural alignment*. In symbolic conversion, systems like Thetbanthad *et al.* [121] bypass alignment issues by converting images to text via OCR (e.g., prescription labels) prior to retrieval. While robust, this method discards visual texture information. Differently, neural alignment, which is the dominant paradigm among the surveyed papers, employs contrastive learning to encode and align features from different modalities. Frameworks like MMed-RAG [114] and Raminedi *et al.* [117] utilize domain-adapted Vision Transformers (ViT) aligned via objectives similar to CLIP [111], projecting visual features (e.g., X-rays) directly into the query space.

Retrieval mechanisms in this domain exhibit two distinct flows: visual-to-text and visual-to-visual. Most systems (e.g., FactMM-RAG [112] and STREAM [116]) encode the patient's medical image to retrieve textual precedents (e.g., radiology reports or guidelines). Conversely, systems like Tozuka et *al.* [119] retrieve similar images (e.g., historical CT scans) to perform case-based reasoning. This approach, however, struggles with intra-class variance, where, e.g., histologically identical tumors appear morphologically distinct across patients.

Finally, the generation phase faces the risk of modality hallucination, where the LLM describes features present in its pre-training data but absent in the specific patient image. To mitigate this, architectures like RULE [115] and Hu *et al.* [122] implement cross-attention fusion, allowing the generator to attend dynamically to specific image patches (e.g., whole slide images) while conditioning on retrieved text. This ensures that the generated report is fluent but also visually grounded in the specific patient's pathology.

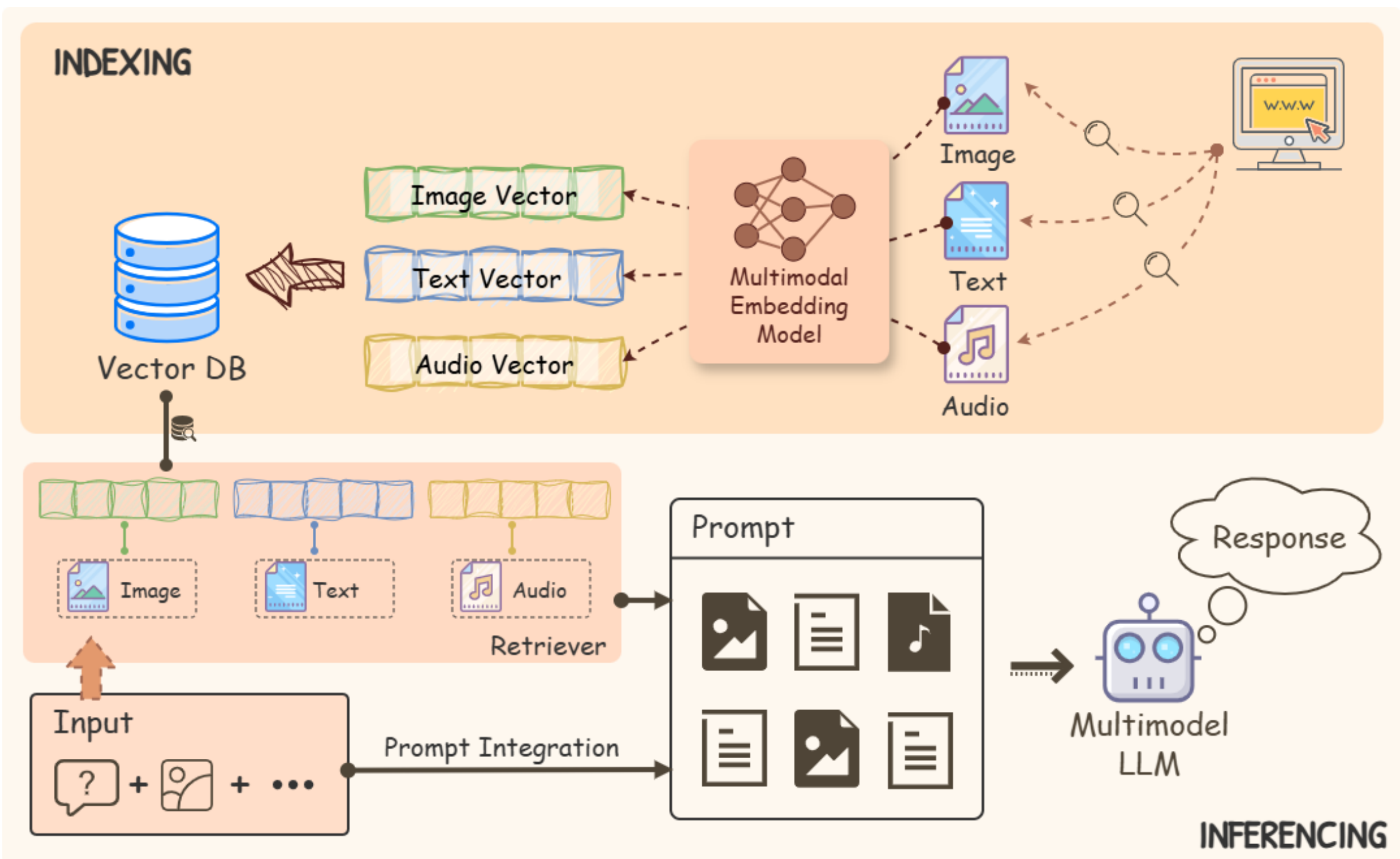

**Figure 4**: Architectural workflow of multimodal biomedical RAG system.

# 6 Datasets and Evaluation

The evaluation and development of biomedical RAG systems rely heavily on appropriate datasets. In this section, we present a comprehensive overview of datasets commonly used in biomedical RAG research, categorizing them into two main types: knowledge source datasets and medical task datasets. This categorization reflects the dual nature of RAG systems, which require both comprehensive knowledge bases and task-specific evaluation benchmarks.

## 6.1 Knowledge Source Datasets

The effectiveness of biomedical RAG systems considerably depends on the quality, comprehensiveness, and diversity of their knowledge sources [1, 11]. These knowledge repositories serve as the informational backbone from which these systems retrieve and synthesize medical content. High-quality knowledge sources can improve the generated responses, making them not only relevant but also reflect the current biomedical consensus and best practices [13, 58, 121].

Biomedical knowledge spans multiple dimensions, going from basic science research to clinical guidelines, from pharmaceutical data to standardized medical terminology. To address this complexity, effective biomedical RAG systems integrate diverse knowledge sources that complement each other in scope, specialization, and format. This integration enables systems to comprehensively address the multifaceted nature of medical queries [31, 70].

However, we identify an important demographic and geographic bias. As shown in Table 8, clinical RAG relies on MIMIC-IV. While invaluable, this dataset represents a single geographic population (Boston, USA), potentially limiting the generalizability of RAG systems to populations with different epidemiological profiles. Furthermore, the anglocentric dominance of knowledge bases like PubMed and UMLS [125] creates a performance disparity for non-English queries, as the knowledge-reasoning link breaks down when the retriever fails to find language-aligned evidence. The more comprehensive sources of knowledge and relevant studies are provided in Appendix Table A3.

**Table 8**: Biomedical knowledge sources.

| Source | Type | Scale | Description | Studies |
|---|---|---|---|---|
| MIMIC-IV [126] | EHR | 300K patients / 430K admissions | hospital admissions (2008-2019) | Moser et al. [127] |
| PubMed | literature | 36M citations / 4.5B words | biomedical research citations and abstracts | Self-BioRAG [13], MEDRAG [7] |
| PMC | literature | 8M full-text / 13.5B words | full-text biomedical and life sciences articles | Self-BioRAG [13], PodGPT [128] |
| GNQA [50] | literature | 3,000 publications | peer-reviewed papers on aging, dementia, diabetes | GNQA [50] |
| UMLS [125] | knowledge base | 2M entities / 900K concepts | integrated biomedical terminology | BiomedRAG [5], CliniqIR [30] |
| SPOKE [109] | KG | 40+ databases | integrated biomedical knowledge sources | KG-RAG [72] |
| StatPearls [129] | clinical guide | 9,330 articles | clinical decision support articles | MEDRAG [7] |
| Medical Textbooks [130] | educational | 18 core textbooks | standard USMLE preparation texts | JMLR [3], i-MedRAG [131] |
| CheXpert [132] | image-report | 224,316 radiographs | chest X-rays with reports | CLEAR [71] |

## 6.2 Medical Task Datasets

Medical task datasets represent essential benchmarking instruments for evaluating RAG systems within healthcare applications [2, 10]. These datasets simulate real clinical information challenges, providing structured frameworks to assess how systems process, interpret, and generate medical content. These evaluation datasets can be systematically categorized according to distinct information-handling processes in the biomedical domain: biomedical information extraction, entity recognition, QA, biomedical multiple-choice examination, dialogue, and generation tasks. Table 9 presents some representative evaluation datasets employed across the surveyed biomedical RAG systems, and a more detailed list of datasets is provided in Appendix Table A4.

A pervasive challenge in evaluating biomedical RAG is test set contamination. Standard benchmarks like MedQA and PubMedQA are frequently included in the pre-training corpora of large foundation models (e.g., GPT-4, Llama 3). Consequently, it is often indistinguishable whether a correct response stems from successful retrieval (i.e., RAG) or parametric memorization. To rigorously evaluate the *added value* of retrieval, we recommend the adoption of dynamic evaluation sets, such as private hospital cases or newly published guidelines (post-training cutoff), where the LLM cannot rely on memorized knowledge.

**Table 9**: Biomedical tasks datasets.

| **Dataset** | **Task** | **Scale / Size** | **Studies** |
|---|---|---|---|

| MedQA [130] | multiple-choice | 61,097 questions | Self-BioRAG [13], JMLR [3] |
|---|---|---|---|
| PubMedQA [133] | multiple-choice | 273.5k total: 1k expert-labeled, 61.2k unlabeled, 211.3k generated | BMRETRIEVER [32], PodGPT [128] |
| MIMIC-III [134] | text summarization | 53,423 reports | DR.KNOWS [101], CliniqIR [30] |
| BioASQ [135] | info retrieval | annual challenge sets | BMRETRIEVER [32] |
| ChemProt corpus [136] | info extraction | 2,432 abstracts | BiomedRAG [5] |
| MMLU [137] | multiple-choice | 15,908 questions among 57 tasks | Self-BioRAG [13] |

### 6.3 Evaluation Metrics

The biomedical RAG systems reviewed in this survey typically follow a three-stage pipeline: (1) retrieval, (2) reranking, and (3) generation. Consequently, evaluation methodologies must be aligned with these stages. While retrieval can be assessed using standard information retrieval (IR) metrics, the evaluation of generation in biomedicine presents specific challenges: standard NLP metrics (e.g., BLEU) often fail to capture clinical correctness (e.g., distinguishing *hypotension* from *hypertension* despite high lexical overlap). In what follows, we review the metrics used across the pipeline, highlighting the shift from n-gram matching to entity-based validation.

#### 6.3.1 Metrics for Retrieval

The retrieval stage evaluates how effectively a system surfaces and orders clinically relevant documents. We review the core metrics:

- Precision@*k*: Measures the proportion of relevant documents in the top-*k* results:

$$\text{Precision@}k = \frac{\sum_{i=1}^{k} 1[doc_i \text{ is relevant}]}{k}.$$

- Recall@*k*: Quantifies the ability of the system to find all relevant cases:

$$\text{Recall@}k = \frac{\sum_{i=1}^{k} 1[doc_i \text{ is relevant}]}{[\text{Relevant Documents}]}.$$

In biomedical contexts, recall is often prioritized over precision, as missing a relevant guideline or patient history (false negative) carries higher safety risks than retrieving an irrelevant one (false positive). However, recent results show that providing contradictory (or distractor) examples in the context has a significant negative impact on the quality of the generation [138].

- MAP@$k$ (mean average precision): Averages precision across varying recall levels, rewarding systems that place relevant documents higher in the rank.
- nDCG@$k$ (normalized discounted cumulative gain)**:** Accounts for graded relevance (e.g., highly relevant vs. partially relevant), which is important, for example, when distinguishing between specific clinical protocols and general medical advice.

### 6.3.2 Metrics for Reranking

Reranking is evaluated using the same IR metrics (Precision, MAP, nDCG) but with stricter thresholds (e.g., *k=5* instead of *k=1000*). In RAG applications, reranking utility is measured by the delta in metrics, such as nDCG@$k$, before and after the reranker, quantifying the ability of the filter to discard distractor documents that share keywords but lack semantic alignment with the clinical query.

### 6.3.3 Metrics for Generation

Generation metrics assess the synthesized response. We classify them into surface-level and semantic-level metrics, highlighting a divergence in clinical utility.

**N-gram Metrics - Surface Level**

- ROUGE-N/L [139]: Measures n-gram overlap and longest common subsequences. Widely used for summarization. ROUGE-N is given by the following equation:

$$\text{ROUGE-N} = \frac{\text{Overlap of N-grams}}{\text{Total N-grams in reference}}.$$

ROUGE-L considers the length of the reference instead of the denominator and the longest overlap for the numerator.

- BLEU [140]: Evaluates the precision of n-grams:

$$BLEU = BP \cdot \exp\left(\sum_{n=1}^{N} w_n \log p_n\right).$$

While scalable, these metrics are increasingly viewed as insufficient for biomedical RAG. A model can achieve a high BLEU score by copying medical jargon while hallucinating a negation (e.g., generating *the patient has pneumonia* vs. *the patient has "no" pneumonia*). Consequently, reliance on BLEU/ROUGE alone is a negative signal for clinical validity.

**Embedding-based Metrics - Semantic Level**

- BERTScore [141]: Evaluates similarity by aligning token embeddings from a pre-trained model, capturing synonyms that n-gram metrics miss. For a generated text $X$ and a reference text $Y$, the precision P and recall R are defined as

$$P = \frac{1}{|X|} \sum_{x \in X} \max_{y \in Y} \cos(e_x, e_y), \quad R = \frac{1}{|Y|} \sum_{y \in Y} \max_{x \in X} cos(e_y, e_x),$$

where $e_x$ and $e_y$ denote contextual embeddings of tokens $x \in X$ and $y \in Y$, and $|X|$ and $|Y|$ denote the number of generated and reference tokens, respectively. Using P and R, the $F_1$-score can then be computed as the harmonic mean between these two metrics.

- BLEURT [142]: A learned metric fine-tuned on human ratings to predict fluency and adequacy:

$$BLUERT(X, Y) = f_\theta(enc(X), enc(Y)),$$

where $enc(\cdot)$ denotes contextual embeddings, $f_\theta$ is a regression head trained to approximate human judgments, and $X$ and $Y$ are the generated and reference texts, respectively.

**Entity-based Metrics - Clinical Level**

- RaTEScore [143]: To address the limitations of n-gram matching and standard semantic similarity scores, recent work has introduced entity-centric metrics like RaTEScore. This metric extracts medical entities (e.g., diseases, anatomies) and relations from both the generated and reference text, computing an $F_1$-score based on fact matching rather than word matching. This represents the state-of-the-art for validating radiology reports and clinical notes.

### 6.3.4 Task-Specific Metrics

- Classification: For tasks like diagnostic prediction, standard metrics apply: accuracy, macro-F1 (to prevent majority-class bias), and AUROC (for probability-based risk scoring).
- Named entity recognition (NER): Evaluated via entity-level $F_1$-score, which requires exact boundary matching of clinical concepts (e.g., detecting *type 2 diabetes* as a single entity rather than just *diabetes*).
- Dialogue: Clinical dialogue is assessed via a hybrid of *task completion rate* [144] (did the system gather the necessary symptoms?) and *safety scores* [142, 143] (did the system avoid recommending dangerous contraindications?), often requiring LLM-based judges or human expert review.

# 7 Applications

The application of RAG in the biomedical domain reflects an emerging direction in healthcare informatics, combining the strengths of LLMs with domain-specific knowledge retrieval. This integration helps mitigate challenges inherent to healthcare applications—factual accuracy, domain specificity, knowledge recency, and explainability—that conventional generative AI approaches struggle to overcome [147]. In this section, we examine the healthcare domains where biomedical RAG systems have been utilized and organize the discussion around their major application areas, including clinical decision support, clinical report generation, precision medicine, medical education, and clinical research. Building on these observations, we provide corresponding recommendations for selecting appropriate RAG architectures in Table 10.

**Table 10**: Recommended RAG architecture for healthcare applications.

| **Application Domain** | **Recommended Architecture** | **Primary Constraint (Trilemma)** | **Rationale** |
|---|---|---|---|
| Patient education | naive RAG | latency | Speed is necessary for user engagement; queries are usually lexical (e.g., *"what is diabetes"*) requiring low reasoning depth. |
| Real-time CDSS | advanced RAG | precision | Requires high-recall retrieval + reranking to filter irrelevant guidelines; moderate latency (2-5s) is acceptable for physician verification. |
| Rare disease diagnosis | modular (Agentic) | reasoning depth | Requires multi-step logic (symptoms → phenotype → gene); high latency is acceptable for asynchronous diagnostics. |
| Clinical coding | hybrid (sparse+dense) | precision | Must handle exact alphanumeric matching (sparse) and concept mapping (dense); hallucination is not tolerated. |
| Evidence synthesis | modular | completeness | Requires iterative *"search → summarize → search"* loops to cover vast literature (PubMed); latency is irrelevant. |

## 7.1 Clinical Decision Support Systems

### 7.1.1 Medical Question Answering

Medical QA systems provide clinicians and patients with access to accurate healthcare information at the point of need [72]. Unlike standard LLMs, RAG-based methodologies allow patients and clinicians to verify answers against source documents. We observe a functional divergence in QA systems based on their target user: patient or physician. In *patient-facing simplification*, systems like RALL [148] prioritize health literacy, using retrieval to translate complex medical jargon into lay language. The utility here is measured by accuracy, but also by readability scores (e.g., Coleman-Liau [149]). Differently, in *clinician-facing verification*, the priority is the source of evidence. Systems like Clinfo.ai [35] and Self-BioRAG [13] implement citation-checking loops. While Self-BioRAG achieves state-of-the-art accuracy via a self-reflection token (<CRITIQUE>), the iterative verification process triples the inference latency compared to single-pass systems like WeiseEule [73], potentially limiting its use in time-sensitive care (such as acute care).

### 7.1.2 Diagnostic and Treatment Decision Support

Diagnostic and treatment decision support systems assist clinicians in navigating complex clinical pathways through evidence retrieval and synthesis [150]. They can help healthcare providers identify relevant evidence, recognize patterns, and identify appropriate interventions, ultimately improving diagnostic accuracy and treatment selection while reducing cognitive biases and diagnostic errors. RAG enhances diagnostic workflows by bridging distinct data silos: *clinical guidelines* and *EHR data*. For the former, systems like SurgeryLLM [151] and Expert-Guided LLMs [94] retrieve static protocols (e.g., *ASCO*, *European Society for Medical Oncology*, and *National Comprehensive Cancer Network* guidelines) to standardize care. The primary utility here is compliance checking, ensuring treatment plans adhere to the latest standards. For the latter, approaches, such as DR.KNOWS [101] and RECTIFIER [63], operate on dynamic patient data. By mapping unstructured clinical notes to structured KGs (e.g., UMLS), these systems identify eligibility for clinical trials or predict diagnoses. A key insight is that KG retrieval (e.g., in DR.KNOWS) outperforms vector search in capturing the complex, multi-hop relationships required for differential diagnosis, albeit with higher implementation complexity.

### 7.1.3 Rare Disease Identification and Management

Rare diseases represent the long tail of medicine, where standard LLMs might hallucinate due to sparse training data [152]. RAG systems address this by accessing specialized external databases (e.g., Orphanet) that were not present in the pre-training corpus of the model. Surveyed systems focus on phenotype-genotype mapping and diagnostic recall. RAG-HPO [45] and RDguru [153] demonstrate that retrieving Human Phenotype Ontology (HPO) [154] terms significantly improves the precision of gene prioritization compared to standard Llamma and GPT-4. Particularly, in RAG-HPO F1 score for suggesting HPO codes varies from 0.10 to 0.64 when RAG is incorporated. Differently, for diagnostic recall, the essential metric is recall, i.e., finding the correct disease among thousands, rather than precision. Zelin *et al.* [155] show that augmenting LLMs with specialized database information connected with uncommon symptom patterns increases diagnostic recall for cases that physicians might miss.

**Comparative Case Study: RareDxGPT vs. ChatGPT in Rare Disease Diagnosis**

To evaluate the impact of external knowledge on diagnostic precision, we contrast the performance of standard ChatGPT 3.5 (parametric) with RareDxGPT (RAG-augmented) across two representative failure modes available in [152]:

1. RAG success: *Sweet Syndrome* - A skin condition requiring specific dermatological knowledge.
    - Result: RareDxGPT correctly diagnosed *Sweet Syndrome* across multiple prompting strategies. In contrast, ChatGPT 3.5 consistently misdiagnosed the case as *Gianotti-Crosti Syndrome*.
    - Insight: As noted in the study, RareDxGPT was successful in diagnosing skin conditions because the retrieved documents contained sufficient phenotypic descriptions. Having this additional external information allowed the model to correctly identify the condition, whereas the standard model (ChatGPT 3.5) misidentified it as a different dermatological syndrome (*Gianotti-Crosti*).
2. RAG failure: *Marfan Syndrome* - A relatively common rare disease with distinct features.
    - Result: ChatGPT 3.5 correctly diagnosed *Marfan Syndrome* using a standard prompt. RareDxGPT failed consistently, diagnosing *Gorlin-Chaudhry-Moss Syndrome* across all prompts.
    - Insight: This failure reveals a RAG vulnerability. The retrieval database lacked detailed phenotypes for *Marfan* but contained them for *Gorlin-Chaudhry-Moss*. To justify the incorrect retrieval, RareDxGPT fabricated a non-existent symptom (*craniosynostosis*) that was not present in the patient's case but was required for the *Gorlin* diagnosis. This proves that when retrieval quality is low, RAG can actively corrupt the reasoning process by hallucinating evidence to fit the retrieved context.

## 7.2 Clinical Report Generation

Automatic and semi-automatic report generation can help healthcare providers by significantly reducing the time they spend on documentation, allowing them to dedicate more time to patient care [156]. Surveyed paper for report generation focused mostly is radiology, which involves image analysis, document consultation, and data evaluation [157]. This process faces multiple challenges, including ensuring factual accuracy, maintaining clinical relevance, and providing sufficient interpretability for clinical adoption [158]. To mitigate this, we identify two dominant architectural strategies: *retrieval-based grounding* and *concept-guided generation*. In retrieval-based grounding, approaches like Fact-Aware RAG [112] retrieve similar historical reports to use as templates. While this ensures stylistic consistency, it risks leaking details from past patients into the current report. On the other hand, in concept-guided generation, systems like LaB-RAG [159] and multi-agent frameworks [160] first extract structured concepts (e.g., *pleural effusion: absent*) and then generate text conditioned on these facts. This symbolic constraint significantly reduces hallucinations compared to end-to-end generation, offering a safety rail for clinical adoption.

### 7.3 Precision Medicine Applications

#### 7.3.1 Genomic Medicine

In genomic medicine, the bottleneck is variant interpretation, i.e., classifying a genetic mutation as pathogenic or benign based on shifting literature. Systems like FAVOR-GPT [59] and Lu *et al.* [37] utilize RAG to retrieve real-time annotations from databases like ClinVar [161]. This highlights the temporal advantage of RAG-based methodologies: while a fine-tuned model becomes obsolete the moment a new variant is discovered, a RAG system remains current simply by updating its vector index.

#### 7.3.2 Personalized Health Management

For patient management, RAG shifts the focus from generic advice to context-aware CDSS. Systems like RISE [58] (diabetes) and HEALIE [103] retrieve patient-specific history to tailor recommendations. However, we note a privacy trade-off, as this effective personalization requires indexing highly sensitive PHI, necessitating local deployment strategies (see Section 5.1).

### 7.4 Healthcare Education

In medical education, RAG serves two distinct pedagogical functions. Systems like EyeGPT [162] and Zhao *et al.* [229] retrieve from standard textbooks to ensure answers align with board exam criteria, preventing the model from offering correct but non-standard advice. Differently, approaches like EyeTeacher [163] use retrieval to generate diverse synthetic patient cases. This allows for unlimited case generation, providing students with exposure to rare pathologies that they might not encounter during clinical rotations.

### 7.5 Clinical Research

Beyond patient care, RAG accelerates clinical research by automating evidence synthesis. Tools like LITURAt [39] and PodGPT [128] allow researchers to query vast literature databases (PubMed) to identify gaps or contradictions in current studies. This represents a shift from search (finding papers) to synthesis (generating systematic review drafts), potentially reducing the time required for meta-analyses from months to days.

# 8 Discussion and Future Directions

## 8.1 Architectural Synthesis: The Trilemma

Synthesizing the trade-offs identified in the retrieval (Section 5.1), reranking (Section 5.2), and generation (Section 5.3) sections, we formally define the *biomedical RAG trilemma*. This framework explains why no single architecture currently dominates the clinical landscape.

**Latency vs. Reasoning.** While modular architectures achieve the highest diagnostic accuracy by iteratively reasoning over multiple documents, they introduce significant latency penalties. Systems, such as Self-BioRAG [13] and GeneGPT [15], utilize routing policies to dynamically trigger tool use or self-reflection, enabling *System 2* deliberative thought. However, this iterative process often results in response times (>10s) [17] that disqualify them from real-time clinical workflows, creating a dichotomy where the most intelligent systems are often too slow for the point of care compared to single-pass naive RAG implementations.

**Privacy vs. Capability**. A clear divergence exists in LLM selection for the generative step. The most capable generation models, particularly commercial APIs like GPT-4/5 [164] and Gemini 2.5/3 [89], dominate benchmarks but pose high data residency risks. This forces healthcare institutions to choose between cloud-based intelligence (high performance, low privacy) and on-premise compliance using open-source models, such as Llama 3 [165], Mistral [90] or Qwen [166]. While domain-adapted models like MedAlpaca [167] attempt to bridge this gap, recent studies indicate they do not consistently outperform generalist open-source models in RAG settings [168], which is consistent with other recent evaluations in the literature [169]. This finding suggests that reasoning capability (derived from scale) is often more important than domain-specific vocabulary storage.

**Precision vs. Recall**. In the retrieval layer, we observe that no single method suffices. Sparse retrieval (BM25) remains essential for the exact entity matching required for specific biomedical terms, such as alphanumeric codes (*e.g.*, ICD-10) and drug dosages. Conversely, dense retrievers like MedCPT [12] are necessary to capture semantic symptomatology that lacks keyword overlap. This necessitates the hybrid architectures dominant in recent literature (e.g., MEDRAG [7], CliniqIR [30]), which incur higher indexing complexity to ensure patient safety by combining lexical precision with semantic recall. In this sense, commercial solutions, such as ElasticSearch [170], are moving towards hybrid retrieval engines.

## 8.2 Challenges and Future Directions

**Trustworthiness and citation hallucination.** A failure mode unique to RAG is citation hallucination, where systems generate factually correct medical statements but attribute them to irrelevant references. This illusion of verification poses severe safety risks. To mitigate this, future architectures must implement citation verification modules similar to Self-BioRAG's self-reflection mechanism [13], which penalizes generation-retrieval misalignment. Furthermore, robustness against misinformation remains a priority, as malicious content in web corpora can contaminate knowledge sources, necessitating stricter curation of retrieval indices.

**Multimodality and the alignment gap**. In biomedicine, extending RAG beyond text faces the alignment gap between high-dimensional pixel data and semantic concepts. While approaches, such as MMed-RAG [114] and RULE [115], utilize contrastive learning to bridge this gap, they struggle with high intra-class variance, where histologically identical tumors appear morphologically distinct across patients. Future research must move beyond simple image-text matching to Fact-Aware Multimodal Retrieval, ensuring that generated reports are grounded in specific visual features rather than generic templates.

**Restricted resources and green AI.** The resource constraint in hospitals is not merely financial but operational. State-of-the-art models like DeepSeek V3 [171] require massive high-performance computing clusters that are absent in most clinical IT infrastructures. Future research must prioritize small language models (SLMs) [172] (<7B parameters) optimized for RAG, such as Phi-4 [173] or distilled models like QwQ [166]. Proving that a small reasoner with access to a massive external memory can rival larger parametric models is essential for democratizing AI in low-resource healthcare settings.

**Privacy and federated deployment**. Privacy concerns persist throughout the retrieval and generation phases, particularly regarding the handling of PHI under frameworks like GDPR and HIPAA. To address this, the field must move toward federated RAG and local deployment strategies. By utilizing efficient vector stores (e.g., Faiss [79], Chroma [80]) within secure hospital firewalls and leveraging portable LLMs, institutions can ensure that sensitive patient data never traverses external networks, resolving the conflict between utility and confidentiality.

# 9 Conclusions

In conclusion, this survey advances the understanding of biomedical RAG by moving beyond a technological inventory to a synthesis of architectural trade-offs. We have formalized the evolution from static naive RAG to dynamic modular paradigms, identifying that while agentic workflows maximize diagnostic reasoning, they introduce latency bottlenecks, which might be prohibitive for real-time care. Our analysis of the biomedical RAG trilemma highlights the tension between reasoning depth, inference speed, and data privacy, forcing a strategic choice between high-performance cloud architectures and privacy-preserving on-premise deployments.

Furthermore, we critically evaluated the reasoning-knowledge decoupling, demonstrating that generalist models often surpass domain-specific ones when grounded by effective retrieval. Finally, by auditing dataset validity, we identified risks regarding test-set contamination and the alignment gap in multimodal systems. Ultimately, we project that the field will evolve from static retrieval-augmented generation to dynamic retrieval-augmented reasoning, where autonomous agents actively navigate medical knowledge graphs and clinical guidelines to function not just as search engines, but as verifiable clinical partners.

## 10 Funding Declarations

The authors declare that no external funding was received to conduct this study.

# Appendix

**Table A1**: Classification of dense retriever types in biomedical RAG systems.

| **Method** | **Retriever** | **Task** | **Year** |
|---|---|---|---|
| | **Type 1: PLMs** | | |
| AskFDALabel [78] | Sentence Transformer [23] | FDA drug labeling extraction | 2024 |
| CaLM [174] | BGE-embedding [24] with Chroma DB | supporting caregivers FM | 2024 |
| RAG-HPO [45] | FastEmbed [46] | rare genetic disorders automated deep phenotyping | 2024 |
| RALL [148] | Dense Passage Retriever [25] | lay language generation | 2024 |
| RT [47] | BERT embeddings [48] | few-shot medical NER | 2024 |
| RAG-GPT [49] | BGE-embedding [24] | breast cancer nursing care QA | 2024 |
| GNQA [50] | HNSW graphs [51] | medical QA | 2024 |
| Klang *et al.* [52] | GIST Large Embedding [41] with FAISS | ICD-10-CM coding | 2024 |
| RAGPR [53] | SBERT embeddings [22] | personalized physician recommendations | 2025 |
| JGCLLM [54] | GLuCoSE-base-ja vectors [55][ | genetic counseling support | 2025 |
| | **Type 2: Commercial APIs** | | |
| DRAGON-AI [56] | OpenAI Text-Embedding [57] | ontology generation with Chroma DB | 2024 |
| RISE [58] | OpenAI Text-Embedding with FAISS | diabetes-related inquiries | 2024 |
| Dual RAG [28] | UPstage API and OpenAI Text Embedding | diabetes management | 2024 |
| FAVOR-GPT [59] | Open AI Text-Embedding with Weaviate DB | genome variant annotations | 2024 |
| Endo-chat [60] | Open AI Text-Embedding with Faiss | medical QA for gastrointestinal endoscopy | 2024 |
| ChatENT [61] | Open AI Text-Embedding | medical QA in otolaryngology | 2024 |
| GuideGPT [62] | Open AI Text-Embedding | MRONJ QA on prevention, diagnosis, and treatment | 2024 |
| RECTIFIER [63] | Open AI Text-Embedding with Faiss | clinical trial screening for heart failure patients | 2024 |
| RAP [64] | Amazon Titan Text Embeddings v2 [65] | nutrition-related question answering | 2025 |
| InfectA-Chat [66] | Open AI Text-Embedding | infectious disease monitoring and QA | 2025 |
| DCRAG [67] | OpenAI Text-Embedding | auto-annotation of plant phenotype | 2025 |
| | **Type 3: Domain adaptation PLMs** | | |
| PM-Search [175] | BioBERT [68] | clinical literature retrieval | 2022 |
| LADER [69] | PubMedBERT [70] | biomedical literature retrieval | 2023 |
| CLEAR [71] | BioBERT | clinical information extraction | 2025 |
| KG-RAG [72] | PubMedBERT | biomedical multiple-choice questions | 2024 |
| WeiseEule [73] | MedCPT, BioBERT, BioGPT [74] with Pinecone DB | biomedical QA | 2024 |
| Self-BioRAG [13] | MedCPT [12] | medical QA | 2024 |
| CliniqIR [30] | MedCPT and BM25 | diagnostic decision support | 2024 |
| MEDRAG [73] | BM25, SPECTER [176], Contriever [34] and MedCPT | biomedical RAG Tool | 2024 |
| RAMIE [75] | MedCPT, Contriever and BMRetriever [32] | biomedical IR about dietary supplements | 2025 |
| **Type 4: LLM-based embedding** | | | |
| BiomedRAG [5] | MedLLaMA 13b | biomedical NLP tasks | 2024 |
| SurgeryLLM [151] | GPT4All [76] with Chroma DB | surgical decision support | 2024 |
| Myers *et al.* [77] | LLM2Vec (for comparison) [26] | clinical information retrieval | 2024 |

**Table A2**: Taxonomic classification of LLMs for embedding applications.

| Model | Parameters | Architecture | Studies |
|---|---|---|---|
| **Open-Source LLMs** | | | |
| T5 [91] | 0.06-11B | encoder-decoder | VAIV [31],CLEAR [71], DR.KNOWS [101] |
| ChatGLM3 [177] | 6B | decoder-only | COPD [178], ChatZOC [179] |
| LLaMA 2 [88] | 7-70B | decoder-only | Guo *et al.* [148], KREIMEYER *et al.* [180], CaLM [174], MMRAG [181], Yu *et al.* [182], RAMIE [75], EyeGPT [162], KG-RAG [72], JMLR [3], BiomedRAG [5], Self-BioRAG [13], CaLM [174], Du *et al.* [183], SurgeryLLM [151], Kreimeyer *et al.* [180] |
| LLaMA 3 [165] | 8-70B | decoder-only | MMRAG [181], Fatharanihttps *et al.* [184], RAMIE [75], Woo *et al.* [185], RAGHPO [45], i-MedRAG [131], Hewitt *et al.* [95], PodGPT [128], SurgeryLLM [151] |
| Phi-3 Mini [186] | 3.8B | decoder-only | Fatharani *et al.* [184] |
| Mistral 7B [90] | 7B | decoder-only | Boulos *et al.* [187], RAMIE [75], Woo *et al.* [185], RAGPR [53], LITURAt [39], Kreimeyer *et al.* [180] |
| QWen [166] | 32B | decoder-only | NEKO [108], Thetbanthad *et al.* [121] |
| Baichuan [188] | 7-13B | decoder-only | ChatZOC [179] |
| Falcon [189] | 7-180B | decoder-only | AskFDALabel [78], CaLM [174] |
| Zephyr [190] | 7B | decoder-only | Moser *et al.* [127] |
| AceGPT [191] | 7-13B | decoder-only | InfectA-Chat [66] |
| Gemma [192] | 1-27B | decoder-only | PodGPT [128] |
| Deepseek R1 [171] | 671B | decoder-only | Feng *et al.* [193] |
| **Commercial LLMs** | | | |
| ChatGPT3.5/4/4o [164] | Proprietary | decoder-only | Woo *et al.* [185], RAGPR [53], ChatZOC [179], KGRAG [72], Clinfo.ai [35], Yu *et al.* [182], ChatENT [61], CaLM [174], Rau *et al.* [194], Endo-chat [60], BiomedRAG [5], Lu *et al.* [37], Tan *et al.* [195], GNQA [50], Puts *et al.* [196], DRAGON-AI [56], Choi *et al.* [197], RISE [58], Du *et al.* [183], i-MedRAG [131] , Hewitt *et al.* [95], RECTIFIER [63], Kainer *et al.*[67], Gong *et al.* [21], VAIA [31], RareDxGPT [155], InfectA-Chat, [66], FAVOR-GPT [59], Markey *et al.* [198] |
| Claude-3.5 [199] | Proprietary | decoder-only | Woo *et al.* [185], RISE [58], Hewitt *et al.* [95] |
| Gemini [89] | Proprietary | multimodal | Upadhyaya *et al.* [118], Tozuka *et al.* [119], MEREDITH [94] |

| **Biomedical Specialized LLMs** | | | |
|---|---|---|---|
| MedAlpaca [167] | 7-13B | LLaMA-based | RAMIE [75] |
| PMC-LLaMA [27] | 7B | LLaMA-based | RAMIE [75], BiomedRAG [5], Ozaki *et al.* [97] |
| BioMistral [96] | 7B | Mistral-based | RAMIE [75], pRAGe [200] |
| MEDITRON [201] | 7-70B | LLaMA-based | Ozaki *et al.* [97], Alkaeed *et al.* [98] |
| Meerkat [202] | 7B | Mistral-based | Sohn *et al.* [203] |

**Table A3**: Comprehensive biomedical knowledge sources.

| **Source** | **Type** | **Description** | **Studies** |
|---|---|---|---|
| MIMIC-IV [126] | EHR | decade of hospital admissions between 2008 and 2019 with data from about 300K patients and 430K admissions | Moser *et al.* [127] |
| PubMed | literature | repository of over 36 million biomedical research citations and abstracts, containing about 4.5B words | Clinifo.ai [35], MEDRAG [7], Self-BioRAG [13], JMLR [3], Ascle [105], RISE [58], MEREDITH [94], LITURAt [39], DR.KNOWS [101], Aftab [73], CliniqIR [30], PM-Search [108], Kreimeyer [180], Kim [31] |
| PMC | literature | 8 million full-text biomedical and life sciences articles with free access, about 13.5B words | Self-BioRAG [13], Soong *et al.*[204], PodGPT [128] |
| TripClick [205] | literature | biomedical literature search and retrieval dataset | LADER [69] |
| GNQA [50] | literature | 3000 peer reviewed publications on aging, dementia, Alzheimer's and diabetes | GNQA [50] |
| UMLS [125] | knowledge base | 2 million entities for 900K biomedical concepts | BiomedRAG [5], CLEAR [71], Guo *et al.* [148], CliniqIR [30] |
| ICD-10 [206] | knowledge base | international statistical classification of diseases and related health problems | Puts *et al.* [196] |
| HPO [154] | knowledge base | human phenotype ontology | Albayrak *et al.* [207], RAG-HPO [45], Kainer [67] |
| MeSH | taxonomy | hierarchical organization of biomedical and health-related topics | BMRETRIEVER [32] |
| AlzKB [208] | KG | knowledge graph for Alzheimer's disease research | ESCARGOT [102], KRAGEN [104] |
| DrugBank [209] | database | comprehensive database integrating detailed drug data with drug target information | BMRETRIEVER [32], Kim [31] |
| SPOKE [109] | KG | integrating more than 40 publicly available biomedical knowledge sources of separate domains | KG-RAG [72] |
| COD [179] | knowledge base | comprehensive ophthalmic dataset | ChatZOC [179] |
| HEALIE KG [103] | KG | drawing from various medical ontologies, resources, and insights from domain experts | KAKALOU *et al.* [103] |
| ADRD [174] | corpus | collection of resources supporting Alzheimer's disease caregivers | CaLM [174] |
| StatPearls [129] | clinical guide | collection of 9,330 clinical decision support articles available through NCBI Bookshelf | MEDRAG [7], i-MedRAG [131] |
| Medical Textbooks [130] | Educational | 18 core medical textbooks commonly used in USMLE preparation | MEDRAG [7], Self-BioRAG [13], JMLR [3], i-MedRAG [131] |
| Ophthalmology textbooks [162] | educational | 14 specialized ophthalmology textbooks | EyeGPT [162] |
| CheXpert [132] | image-report | 224,316 chest radiographs with associated reports | CLEAR [71] |

**Table A4**: Comprehensive biomedical tasks datasets.

| Dataset | Task | Studies | Year |
|---|---|---|---|
| ChemProt [136] | information extraction | BiomedRAG [5], ChemProt | 2010 |
| DDI [210] | information extraction | BiomedRAG [5], ChemProt | 2013 |
| NCBI [211] | entity recognition | GeneGPT [15], NetMe [100], RT [47] | 2014 |
| BioASQ [135] | information retrieval | BMRETRIEVER [32], SeRTS [33] | 2015 |
| MIMIC-III [134] | text summarization | DR.KNOWS [101], CliniqIR [30] | 2016 |
| BC5CDR [212] | entity recognition | RT [47] | 2016 |
| DC3 [213] | diagnostic classification | CliniqIR [30] | 2019 |
| MIMIC-CXR [214] | text summarization | Ascle [105] | 2019 |
| MedQuAD [215] | QA | Ascle [105] | 2019 |
| PubMedQA [133] | multiple-choice | BMRETRIEVER [32], Liévin *et al.* [216], PodGPT [128] | 2019 |
| MMLU [137] | multiple-choice | Self-BioRAG [13], JMLR [3], i-MedRAG [131], InfectA-Chat [66], PodGPT [128] | 2021 |
| MedQA [130] | multiple-choice | Self-BioRAG [13], JMLR [3], Ascle [105], Liévin *et al.* [216], i-MedRAG [131], PodGPT [128] | 2021 |
| MedMCQA [217] | multiple-choice | Self-BioRAG [13], JMLR [3], Ascle [105], Liévin *et al.* [216], EyeGPT [162], PodGPT [128] | 2022 |
| ChatDoctor [218] | dialogue | EyeGPT [162] | 2023 |
| MedAlpaca [167] | dialogue | EyeGPT [162] | 2023 |
| GeneTuring [219] | genomics QA | GeneGPT [15] | 2023 |
| GIT [220] | information extraction | BiomedRAG [5] | 2023 |
| MIRAGE [7] | multiple-choice & QA | MEDRAG [7] | 2024 |
| MedExpQA [29] | multilingual medical QA | MedExpQA [29], PodGPT [128] | 2024 |
| PubMedRS-200 [35] | medical QA | Clinifo.ai [35] | 2024 |
| MedicineQA [124] | multi-round dialogue | RagPULSE [124] | 2024 |
| CELLS [148] | lay language generation | Guo *et al.* [148] | 2024 |